\documentclass[12pt]{article}
\usepackage{simplemargins, amsmath, array, delarray, emlines2, epsfig,
newlfont}
\setallmargins{1in}
\begin{document}

\thispagestyle{empty}
\begin{flushright}
hep-th/0303067
\end{flushright}
\vskip 3.5cm

\begin{center}\LARGE
{\bf A Quantum Bousso Bound}
\end{center}
\vskip 1.0cm
\begin{center}
{\large  Andrew Strominger
and David Thompson}

\vskip 0.5 cm
{\it Jefferson Physical Laboratory\\
Harvard University\\
Cambridge, MA  02138}
\end{center}

\vskip 2cm


\begin{abstract}

The Bousso bound requires that one quarter the area of a closed
codimension two spacelike surface exceeds the entropy flux across
a certain lightsheet terminating on the surface. The bound can be
violated by quantum effects such as Hawking radiation. It is
proposed that at the quantum level the bound be modified by adding
to the area the quantum entanglement entropy across the surface.
The validity of this quantum Bousso bound is proven in a
two-dimensional large N dilaton gravity theory.

\end{abstract}

\vfill
\newpage
\renewcommand{\baselinestretch}{1.4}

\section{Introduction}

The generalized second law of thermodynamics (GSL) \cite{bekenstein}
roughly speaking states that one quarter the area of black hole horizons
plus the entropy outside the horizons is nondecreasing. This law was
formulated in an attempt to repair inconsistencies in the ordinary second
law in the presence of black holes.  There is no precise general
statement, let alone proof, of the GSL, but it has been demonstrated in a
compelling variety of special circumstances. It indicates a deep
connection between geometry, thermodynamics and quantum mechanics which we
have yet to fathom. The holographic principle \cite{thooft93,
susskind95}, which also has no precise general statement, endeavors to
elevate and extend the GSL to contexts not necessarily involving black
holes. In \cite{bousso99}, a mathematically precise modification of the
GSL/holographic principle was proposed that is applicable to null surfaces 
which are {\em not} horizons \cite{bousso02}. This proposed ``Bousso 
bound", along with a generalization stated therein, was proven, subject 
to certain conditions, in a classical limit by Flanagan, Marolf, and Wald 
\cite{flanagan99}.

The Bousso bound, as stated, can be violated by quantum effects
\cite{lowe99}. Mathematically, the proofs of the bound rely on the local
positivity of the stress tensor which does not hold in the quantum world.
Physically, the bound does not account for entropy carried by Hawking
radiation. In this paper, we propose that, at a semiclassical level, the
bound can be restored by adding to one quarter the surface area the
entanglement entropy across the surface. We will make this statement fully
precise, and then prove it, in a two-dimensional model of large N dilaton
gravity.

This paper is organized as follows.  We begin by reviewing Bousso's
covariant entropy bound in section 2.  We will review the lightsheet
construction in general D-dimensional spacetime, although our main
interest in the remainder of the paper will be four and two dimensions. In
section 3, we will discuss how Bousso's bound can be violated in the
presence of semiclassical effects, like Hawking radiation.  This will
motivate us to propose a ``quantum Bousso bound'' in section 4.  By
assuming an adiabaticity condition on the entropy flux, we will show in
section 5 that the classical Bousso bound can be proven in four and two
dimensions.  In section 6, we extend the analysis to the two-dimensional
RST quantization \cite{rst} of the CGHS model \cite{cghs} which includes
semiclassical Hawking radiation and its backreaction.  We will show that
the quantum Bousso bound holds in this gravitational theory.

\section{Review of the classical Bousso bounds}

The Bousso bound asserts that, subject to certain assumptions, the
entropy of matter that passes through certain lightsheets
associated with a given codimension two spatial surface in
spacetime is bounded by the area of that surface \cite{bousso99}.

This entropy bound provides a covariant recipe for associating a
geometric entropy with any spatial surface $B$ that is codimension
two in the spacetime.  At each point of $B$, there are four null
directions orthogonal to $B$.  These four null directions single
out four unique null geodesics emanating from each point of $B$:
two future-directed and two past-directed. Without loss of
generality, we choose an affine parameter $\lambda$ on each of
these curves such that $\lambda$ equals zero on $B$ and increases
positively as the geodesic is followed away from $B$.

Along each of the four geodesics, labelled by $i$, an expansion parameter
$\theta_i(\lambda) = \nabla_a \left(\frac{d}{d\lambda}\right)^a$ can be
defined.  If we note that each of the
future-directed geodesics is simply the extension of one of the
past-directed geodesics, then the following relations between the
expansion parameters becomes clear:  $\theta_1(0)  = -\theta_3(0)$,
$\theta_2(0) = -\theta_4(0)$.  Therefore, at least two of the four
geodesics will begin with a nonpositive expansion.  A ``lightsheet'' is
a codimension one surface generated by following exactly one non-expanding
geodesic from each point of $B$.  Each geodesic is followed until one
of the following occurs on it:
\begin{itemize}
\item The expansion parameter becomes positive, $\theta > 0$,
\item A spacetime singularity is reached.
\end{itemize}
Note that, in spacetime dimensions greater than two, there are an infinite
number of possible lightsheets to choose from since, for each point on
$B$, there are at least two contracting null geodesics from which to
choose.

The original Bousso bound conjectures that Nature obeys the
following inequality:
\begin{equation}
\text{\bf Entropy passing through any lightsheet of } B \; \leq \;
\frac{1}{4} \left(\text{ \bf Area of } B\right) \, .
\end{equation}
In order to make this statement precise, we must clarify what we
mean by the entropy that passes through a lightsheet.  In general,
this is ambiguous because entropy is not a local concept. However,
there is a thermodynamic limit in which the entropy is
well-approximated by the flux of a four-vector
$s^a$.  As discussed by Flanagan, Marolf, and Wald (FMW) in
\cite{flanagan99}, this thermodynamic limit is satisfied under the
entropy condition that we will use in sections 5 and 6.
The Bousso bound as  formulated so far pertains mainly to this limit.

To find the entropy flux that passes through the lightsheet, we must
project $s^a$ onto $k^b$, the unique future-directed normal to the
lightsheet.  Up to a sign, $k$ is $d/d\lambda$ since $d/d\lambda$ is null
and orthogonal to all other lightsheet tangent vectors by construction.
In order to keep $k^a$
future-directed, we choose $k^a = \left(\frac{d}{d\lambda}\right)^a$ if
the lightsheet is future-directed, and $k^a =
-\left(\frac{d}{d\lambda}\right)^a$ if the lightsheet is past-directed.
Since we use the mostly-positive metric signature, the entropy flux
through any point of the lightsheet is
\begin{equation}
s \equiv -k_a s^a\, .
\end{equation}

In the
language of entropy flux, the entropy bound becomes
\begin{equation}
\int_{L(B)} s \; \leq \; \frac{1}{4} (\text{Area of } B)\, ,
\end{equation}
where $L(B)$ denotes the lightsheet of $B$. However, there is a
generalized Bousso bound \cite{flanagan99} in which  the lightsheet is
prematurely terminated on a spatial surface $B'$.  It is clear
that the integral of $s$ over this terminated lightsheet equals
the integral over the full lightsheet of $B$ minus the integral
over the full lightsheet of $B'$. Assuming that $s$ is everywhere
positive, Bousso's original entropy bound tells us that
\begin{equation}
\int_{L(B-B')}s \; \leq \; \int_{L(B)} s \; \leq \; \frac{1}{4}
(\text{Area of } B)\,
,
\end{equation}
where $L(B-B')$ denotes the lightsheet of $B$ terminated on $B'$.
In this paper, we will be interested in the generalized Bousso
bound, first proposed by FMW \cite{flanagan99},
which imposes the much stronger bound on the terminated
lightsheet:
\begin{equation}
\int_{L(B-B')} s \; \leq \; \frac{1}{4} \left(A(B) - A(B')\right)\, .
\label{eqclassbound}
\end{equation}
This has been proven under suitable assumptions by FMW \cite{flanagan99}.
Note that this generalized entropy bound directly implies Bousso's
original entropy bound.

\section{Semiclassical violations}

The entropy bounds so far pertain largely to the classical regime.
When quantum effects are included, even at the semiclassical
level, we expect that the bounds must be somehow modified to
account for the entropy carried by Hawking radiation.
Mathematically, the proofs \cite{flanagan99} are not applicable
because quantum effects violate the positive energy condition.

The classical proofs hinge on the focussing theorem of classical
general relativity. The focussing theorem, in turn, derives from
the Raychaudhuri equation and the null energy condition.  The
Raychaudhuri equation provides a differential equation for the
expansion parameter along a null geodesic \cite{mtw}:
\begin{equation}
\frac{d\theta}{d\lambda} = -\frac{1}{D-2} \theta^2 - \sigma_{ab}
\sigma^{ab} + \omega_{ab} \omega^{ab} - 8 \pi T_{ab} k^a k^b\, ,
\label{eqray}
\end{equation}
where $\sigma_{ab}$ is the shear tensor and $\omega_{ab}$ is the twist
tensor.
For a family of null geodesics that start off orthogonal to a spatial
surface, such as the case for a lightsheet, the twist tensor is
zero.  Finally, if we assume that the null energy condition holds, then
the last term is negative.  The null energy condition postulates that
$T_{ab}k^a k^b$ is nonnegative for all null vectors $k^a$.  As a result,
we
find that the expansion parameter satisfies the inequality
\begin{equation}
\frac{d\theta}{d\lambda} \leq -\frac{1}{D-2} \theta^2\, .
\end{equation}
This gives us the focussing theorem: If the expansion parameter takes the
negative value $\theta_0$ along a null
geodesic of the lightsheet, then that geodesic will reach a caustic (i.e.,
$\theta \rightarrow -\infty$) within the finite affine time
$\Delta \lambda \leq \frac{D-2}{|\theta_0 |}$.

So long as energy is required to produce entropy, the focussing
theorem ensures that the presence of entropy will cause the
lightsheet to reach a caustic and, therefore, terminate.  The more
entropy we try to pass through the lightsheet, the faster the
lightsheet terminates.  This gives a compelling argument for why
only a finite, bounded amount of entropy could be passed through
the lightsheet.  According to the Bousso bound, this upper bound
is precisely one quarter the area of the generating surface.

In practice, the covariant entropy bound can be violated in the presence
of matter with negative energy. By mixing positive-energy matter and
negative-energy matter, a system with zero energy can be made to carry an
arbitrary amount of entropy.  Again, the entropy passing through any given
lightsheet could be increased arbitrarily.  At the classical level, we
could simply demand that the energy-momentum tensor obeys the null energy
energy condition. This is the weakest of all the most common energy
conditions and, as can be seen from (\ref{eqray}), is the one needed for
the focussing theorem, and thus to make the Bousso bound plausible.

However, the Bousso bound is in serious trouble once we include
quantum effects.  We know that none of the local energy conditions
can hold even at first order in $\hbar$. In particular, the
phenomenon of Hawking radiation violates the null energy condition
near the horizon of black holes.  This allows for violations of
the focussing theorem.  This violation can be seen most clearly
for future-directed, outgoing null geodesics that hover for a
while in between the event horizon and apparent horizon of an evaporating
black hole.  The apparent horizon is the boundary of the region of
trapped surfaces, so the congruence of null geodesics are
contracting inside the apparent horizon.  However, as the black
hole evaporates, the apparent horizon follows a timelike
trajectory towards the event horizon. The null geodesic could then
leave the apparent horizon and begin expanding, in violation of
the focussing theorem.

Furthermore, in \cite{lowe99}, Lowe constructs a related
counterexample to the covariant entropy bound in the presence of a
critically illuminated black hole. Critical illumination is the
process in which matter is thrown into a black hole at exactly the
same rate as energy is Hawking radiated away. In this scenario,
the apparent horizon follows a null trajectory.  If we pick the
apparent horizon to be the generating surface for a lightsheet,
then the lightsheet will coincide with the apparent horizon as
long as we continue to critically illuminate the black hole.  By
critically illuminating the black hole sufficiently long, we can
pass an arbitrary amount of matter through the lightsheet.  In
this way, the entropy of the matter passing through the lightsheet
can be made larger than the area of the apparent horizon, thus
violating the entropy bound.

Hence, the original Bousso bound only has a chance of holding in the
classical regime. Once we include one-loop quantum effects, such
as Hawking radiation, the bound fails.  In the remaining sections,
we propose a modification of the Bousso bound which may hold in
the semiclassical regime.

\section{A quantum Bousso bound}

The generalized Bousso bound, when specialized to black hole
horizons, is equivalent to a classical limit of the generalized
second law of thermodynamics (GSL). To see this, note that the portion
of the event horizon lying between any two times consitutes a
lightsheet. Since all matter falling into the black hole between
those two times must pass through this lightsheet, the generalized
entropy bound gives us the same information as the GSL.  In
particular, we learn that
\begin{equation}
\frac{1}{4} \Delta A_{\text{EH}} \geq \Delta S_{\text{m}}\, ,
\label{eqgsl}
\end{equation}
where $\Delta A_{\text{EH}}$ is the change in event horizon area,
and $\Delta S_{\text{m}}$ is the entropy of the matter that fell
in.

When quantum effects are included, the form (\ref{eqgsl}) of the
generalized second law is no longer valid. The quantum GSL states,
roughly speaking,  that the total entropy outside the black hole
plus one quarter the area of the horizon (either event or apparent
depending on the formulation) is non-decreasing. The entropy
outside the black hole receives an important contribution from
Hawking radiation. Therefore, we must augment the left hand side by
the entropy of the Hawking radiation:
\begin{equation}
\frac{1}{4} \Delta A_{\text{H}}+ \Delta S_{\text{hr}}\geq \Delta
S_{\text{m}}\,
\label{eqmgsl}
\end{equation}
In general, we do not know how to formulate, let alone prove, an
exact form of the GSL in a full quantum theory of gravity. However,
approximations to it have been formulated and demonstrated in a
wide variety of circumstances \cite{wald99}. The $\Delta S_{\text{hr}}$
term is crucial in these demonstrations, without which counterexamples may 
be easily found.

Since the GSL requires an additional term at the quantum level,
and the GSL is a special case of the generalized Buosso bound, we
should certainly expect that the Bousso bound will receive related
quantum corrections. These corrections should reduce to $\Delta
S_{\text{hr}}$ when the lightsheets are taken to be portions of event
horizons. The problem is to precisely formulate the nature of
these corrections.

In this context it is useful to think of the  entropy in Hawking
radiation as entanglement entropy. Evolution of the quantum fields
on a fixed black hole geometry is a manifestly unitary process
prior to singularity formation. Nevertheless entropy is created
outside the black hole because the outgoing Hawking quanta are
correlated with those that fall behind the horizon. When a region
of space $U$ is unobservable, we should trace the quantum state
$\psi$ over the modes in the unobservable region to obtain the
observable  density matrix $\rho$,
\begin{equation}
\rho = \text{tr}_U\, | \psi \rangle \langle \psi |\, .
\end{equation}
Since the full state is in principle not available to the
observer, there is a de facto loss of information that can be
characterized by the entanglement entropy
\begin{equation}
S_{\text{ent}} = -\text{tr}\, \rho \log \rho\, .
\end{equation}
In general, this expression has divergences and requires further
definition, which will be given below for the case of two
dimensions.\footnote{UV divergences in this expression are
absorbed by the renormalization of Newton's constant
\cite{susskind94}.} Choosing $U$ to be the region behind the
horizon, we can therefore formally identify
\begin{equation} \Delta S_{\text{hr}}=\Delta S_{\text{ent}}.
\label{fghr}
\end{equation}
This motivates a natural guess for quantum corrections to the
Bousso bound when the initial and final surfaces are closed. One
should add to the area the entanglement entropy across the
surface.  Applying this modification to the classical Bousso bound
(\ref{eqclassbound}) results in a quantum Bousso bound of the
form:
\begin{equation}
\int_{L(B-B')} s \; \leq \; \frac{1}{4} A(B) + S_{\text{ent}}(B) -
\frac{1}{4} A(B') - S_{\text{ent}}(B') \, .
\end{equation}

Since we can not presently hope to solve this problem or even
define this quantum bound in exact quantum gravity, in order to go
further we need to identify a small expansion parameter for
approximating the exact theory. A useful parameter, which
systematically captures the quantum corrections of Hawking
radiation, is provided by $\frac{1}{N}$, where $N$ is the number
of matter fields and $G_NN$ is held fixed \cite{cghs}. In
\cite{fpst} it was shown in the two-dimensional RST model of black
hole evaporation that the (suitably defined) GSL, incorporating
the Hawking radiation as in (\ref{eqmgsl}), is valid.  One might
hope that a similar incorporation can save the Bousso bound.

In the process of the investigations in \cite{fpst} it emerged
that the sum $A+4S_{\text{ent}}\equiv A_{\text{qu}}$ arises naturally in
the theory as a kind of quantum-corrected area. In this paper, we
propose that the required leading $\frac{1}{N}$ semiclassical
correction to the generalized Bousso bound simply involves the
replacement of the classical area with this quantum corrected
area. A precise version of this statement will be formulated and
proved in the RST model in section 6.

\section{Proving the classical Bousso bounds}

In this section we reproduce proofs of classical Bousso bounds. We
first give a proof due to Bousso, Flanagan, and Marolf of the generalized
Bousso bound in four dimensions \cite{rbef}.\footnote{We thank
Raphael Bousso and Eanna Flanagan for explaining this
proof prior to publication.} This simplified proof follows from conditions on
the initial entropy flux and an adiabaticity condition on the rate
of change of the entropy flux which differ somewhat from the
conditions assumed in \cite{flanagan99}.  We then describe a two
dimensional version of the proof obtained by spherical reduction.
A small modification of this gives a proof of the generalized
Bousso bound in the classical CGHS model \cite{cghs}, which is
then transcribed into Kruskal gauge for later convenience. The
inclusion of quantum effects in the latter will be the subject of
the next section.

\subsection{Simplified proof in four dimensions}

Following \cite{flanagan99}, the integral of the entropy flux $s$
over the lightsheet can be written as
\begin{equation}
\int_{L(B-B')} s \; = \; \int_B d^{2}x \sqrt{h(x)} \int_0^1 d\lambda\,
s(x, \lambda) \mathcal{A}(x, \lambda)\, .
\label{fmweq}
\end{equation}
In this expression, we have chosen a coordinate system $(x^1, x^2)$ on the
spatial surface $B$, $h(x)$ is the determinant of the
induced metric on $B$, and the affine parameter on each null geodesic of
the lightsheet has been normalized so that $\lambda = 1$ is when the
geodesic reaches $B'$.  The function $\mathcal{A}(x, \lambda)$ is the
area decrease factor for the geodesic that begins at the point $x$ on
$B$.  In terms of $\theta$, it is given by
\begin{equation}
\mathcal{A} \equiv \exp \left[ \int_0^\lambda d\tilde{\lambda}
\, \theta(\tilde{\lambda}) \right]\, .
\end{equation}
The physical intuition for equation (\ref{fmweq}) is simple.  As we
parallel propagate a small coordinate patch of area $d^2x \sqrt{h(x)}$
from the point $(x, 0)$ on $B$ to the point $(x,\lambda)$ on the
lightsheet, the area contracts to $d^2x \sqrt{h(x)}
\mathcal{A}(x,\lambda)$.  The proper three-dimensional volume of an
infinitesimal cube of the lightsheet is $d^2 x\, d\lambda
\sqrt{h(x)}
\mathcal{A}(x,\lambda)$, and this volume times $s(x,\lambda)$ gives the
entropy flux passing through that cube.
In order to prove the generalized entropy bound, it is
sufficient to prove that
\begin{equation}
\int_0^1 d\lambda\, s(\lambda) \, \mathcal{A}(\lambda) \leq \frac{1}{4}
(1-\mathcal{A}(1))
\end{equation}
for each of the geodesics that comprise the lightsheet.

Using a mostly positive metric signature, the assumed entropy
conditions are

\begin{description}
\item[i. ] $s' \leq 2\pi \, T_{ab} k^a k^b$
\item[ii. ] $s(0) \leq - \frac{1}{4} \,
\mathcal{A}'(0)$,
\end{description}
where we use the notation, both here and henceforth, that primes
denote differentiation with respect to the affine parameter
$\lambda$. Condition \textbf{i} is very similar to one of the
conditions in \cite{flanagan99}. It can be interpreted as the
requirement that the rate of change of the entropy flux is less
than the energy flux, which is a necessary condition for the thermodynamic
approximation to hold.  Condition \textbf{ii} requires only that
the covariant entropy bound is not violated infinitesimally at the
beginning of the lightsheet.
Since the square root of $\mathcal{A}$ routinely appears in calculations,
we borrow the notation of FMW and define
\begin{equation}
G \equiv \sqrt{\mathcal{A}}\, .
\end{equation}

From the Raychaudhuri equation, we have that
\begin{equation}
T_{ab} k^a k^b = -\frac{1}{4\pi} \frac{G''(\lambda)}{G(\lambda)} -
\frac{1}{8\pi} \sigma_{ab} \sigma^{ab} \leq -\frac{1}{4\pi}
\frac{G''(\lambda)}{G(\lambda)}\, ,
\label{eqray2}
\end{equation}
where $\sigma_{ab}$ is the shear tensor, and the inequality follows from
the fact that $\sigma_{ab} \sigma^{ab} \geq 0$ always.
Now we see that
\begin{eqnarray*}
s(\lambda) &=& \int_0^\lambda d\tilde{\lambda}\, s'(\tilde{\lambda}) +
s(0)\\
(i) \quad &\leq& 2\pi \int_0^\lambda d\tilde{\lambda} T_{ab} k^a k^b +
s(0)\\
(eom) \quad &\leq& 2\pi \int_0^\lambda d\tilde{\lambda} \left(
-\frac{1}{4\pi}
\frac{G''(\tilde{\lambda})}{G(\tilde{\lambda})} \right) + s(0)\\
&=& \frac{1}{2} \left( \frac{G'(0)}{G(0)} -
\frac{G'(\lambda)}{G(\lambda)} \right) - \frac{1}{2} \int_0^\lambda
d\tilde{\lambda} \frac{G'(\tilde{\lambda})^2}{G(\tilde{\lambda})^2} +
s(0)\\
(ii) \quad &\leq& -\frac{1}{2} \frac{G'(\lambda)}{G(\lambda)} -
\frac{1}{2} \int_0^\lambda
d\tilde{\lambda} \frac{G'(\tilde{\lambda})^2}{G(\tilde{\lambda})^2}\\
&\leq& -\frac{1}{2} \frac{G'(\lambda)}{G(\lambda)}\, .
\end{eqnarray*}
Consequently,
\begin{equation}
\int_0^1 d\lambda\, s(\lambda) G(\lambda)^2 \leq -\frac{1}{2} \int_0^1
d\lambda\, G(\lambda) G'(\lambda) = \frac{1}{4} \left( \mathcal{A}(0) -
\mathcal{A}(1) \right)\, .
\end{equation}
We have shown that, given our entropy conditions, the entropy passing
through a lightsheet is bounded by one quarter the difference in area of
the two bounding spatial surfaces.  This is precisely the statement of the
generalized Bousso bound.

It is interesting to note that nowhere in the proof did we need to use the
contracting lightsheet condition.  The only indication that we should
choose a contracting lightsheet comes from the boundary condition
\textbf{ii}.  We see from condition \textbf{ii} that, in order to allow a
positive, future-directed entropy flux, the derivative of $\mathcal{A}$
must be negative.  If the lightsheet were expanding at $\lambda = 0$, then
a timelike entropy flux would have to be past-directed at $\lambda = 0$.

Note also that Bousso's entropy bound can be saturated only if $G' = 0$
for all $\lambda$.  In light of the Raychaudhuri equation (\ref{eqray2}),
we see that $T_{ab}$ and the shear $\sigma_{ab}$ must be zero everywhere
along the lightsheet in order for $G'$ to remain zero.  The bound can be
saturated only in this most trivial scenario.  This will not be the case
for other gravitational theories we will study, such as the CGHS dilaton
model, where saturation of the bound can occur in the presence of matter.

\subsection{Spherical reduction}

Our goal is to study the entropy bound in two dimensional models where
our semiclassical analysis will be greatly simplified.
As a guide to what phenomenological conditions we should be using
in 2D models, we will first rederive the previous proof for the
purely spherical sector of 4D Einstein-Hilbert gravity.

We begin with the 4D Einstein-Hilbert action coupled to some matter
Lagrangian density, $\mathcal{L}_m$:
\begin{equation}
\int d^4 x \sqrt{-g^{(4)}} \left(\frac{R^{(4)}}{16 \pi} +
\mathcal{L}^{(4)}_m
\right)\, .
\end{equation}
Writing the four-dimensional metric as
\begin{equation}
(ds^2)^{(4)} = g_{\mu\nu}(x) dx^\mu dx^\nu +
e^{-2\phi(x)} (d\vartheta^2 +
\sin^2 \vartheta \, d\varphi^2)\, \quad \mu,\nu \in \{0,1\}\, ,
\end{equation}
and integrating over the angular coordinates we find the action is reduced
to
\begin{equation}
\int d^2x \sqrt{-g} \left[ e^{-2\phi} \left( \frac{1}{4} R + \frac{1}{2}
g^{\mu\nu}\partial_\mu\phi\partial_\nu\phi + \frac{1}{2}e^{2\phi}
\right) + \mathcal{L}_m \right]\, .
\end{equation}
Here, the 2D matter
Lagrangian density $\mathcal{L}_m$ is related to $\mathcal{L}^{(4)}_m$ by
\begin{equation}
\mathcal{L}_m = 4\pi e^{-2\phi} \mathcal{L}^{(4)}_m\, .
\end{equation}
From the equations of motion, we conclude that
\begin{equation}
k^a k^b T_{ab} = - e^{-\phi} k^a k^b \nabla_a \nabla_b
e^{-\phi}\, ,
\end{equation}
whenever $k$ is a null vector.  In this expression, $T$ is the
energy-momentum tensor for $\mathcal{L}_m$, not $\mathcal{L}_m^{(4)}$.

It is clear from the four-dimensional metric that
the classical ``area'' of a point in the 2D model is $A_{\text{cl}} =
4\pi e^{-2\phi}$.  However, had we only been given the
action, we could identify the ``area'' of a
point as being proportional to the factor multiplying the Ricci scalar in
the Lagrange density.  If
that were not convincing enough, we could study the thermodynamics of a
black hole solution of the two-dimensional model.  In particular, we would
first determine the mass of a stationary black hole solution and then
compute the temperature of the Hawking radiation on this geometry
(neglecting backreaction).  Integrating the thermodynamic identity $dS =
dM / T$ and identifying $S$ as one-quarter the area of the event horizon,
we arrive at an expression for the area of the event horizon in terms of
the local values of the various fields there.  We then designate this
function of local fields as the expression that gives us the ``area'' of
any point in the two-dimensional space.

Deriving the two-dimensional entropy conditions is a simple matter of
rewriting the four-dimensional conditions in terms of two-dimensional
tensors.  For example, we replace $T^{(4)}_{ab}$ with
$\frac{1}{4\pi}e^{2\phi} T_{ab}$.  We are also interested in the
two-dimensional entropy flux $s_a$ which is related to the
four-dimensional entropy flux $s_a^{(4)}$ by $s_a^{(4)} =
\frac{1}{4\pi}e^{2\phi} s_a$.  This relation is a simple consequence
of the fact that the 2D flux at a point equals the 4D flux up to an
overall factor of the area of the corresponding $S^2$.  Replacing 4D
tensors with 2D tensors, we arrive at the following entropy conditions:
\begin{description}
\item[i. ] $e^{-2\phi} (s\, e^{2\phi})' \leq 2\pi T_{ab}\, k^a k^b$
\item[ii. ] $s(0) \leq - \frac{1}{4} A_{\text{cl}}' (0)$
\end{description}
Note that we continue to use $s \equiv -k^a s_a$ and
primes denoting $d/d\lambda$.
Putting it all together, the derivation of the entropy bound goes
through in the same way as it did in the 4D case.  In detail, we find
\begin{eqnarray*}
s(\lambda) &=& e^{-2\phi(\lambda)} \int_0^\lambda d\tilde{\lambda}
\left(s(\tilde{\lambda}) e^{2\phi(\tilde{\lambda})}\right)' +
e^{-2\phi(\lambda)} s(0) e^{2\phi(0)}\\
(i) \quad
&\leq& e^{-2\phi(\lambda)} \int_0^\lambda d\tilde{\lambda}\,
2\pi \, e^{2\phi(\tilde{\lambda})} k^a k^b T_{ab} (\tilde{\lambda}) +
e^{-2\phi(\lambda)} s(0) e^{2\phi(0)}\\
(eom) \quad &=& -2\pi e^{-2\phi(\lambda)} \int^\lambda_0 d\tilde{\lambda}
\left(e^{-\phi(\tilde{\lambda})} \right)'' e^{\phi(\tilde{\lambda})} +
e^{-2\phi(\lambda)} s(0) e^{2\phi(0)}\\
&=& -2\pi e^{-2\phi(\lambda)} \left( -\phi'(\lambda) + \phi'(0)
\right) - 2\pi e^{-2\phi(\lambda)} \int_0^\lambda
d\tilde{\lambda} \left( \phi'(\tilde{\lambda}) \right)^2
+ e^{-2\phi(\lambda)} s(0) e^{2\phi(0)}\\
(ii) \quad &\leq& -\pi \left( e^{-2\phi(\lambda)}
\right)' - 2\pi e^{-2\phi(\lambda)}
\int_0^\lambda d\tilde{\lambda}
\left( \phi'(\tilde{\lambda}) \right)^2 \\
&\leq& -\pi \left( e^{-2\phi(\lambda)} \right)'
\end{eqnarray*}
Therefore,
\begin{equation}
\int_0^1 d\lambda\, s(\lambda) \leq \pi \left( e^{-2\phi(0)} -
e^{-2\phi(1)} \right) = \frac{1}{4} \left(A_{\text{cl}}(0) -
A_{\text{cl}}(1)\right) \, ,
\end{equation}
which is exactly the 4D entropy bound, only derived from the 2D
perspective.

\subsection{CGHS}

Although we have derived an entropy bound in a 2D model using
2D entropy conditions, we were guaranteed success since we had spherically
reduced a successful 4D proof.  We now attempt to apply the same entropy
conditions to another 2D dilaton gravity model, namely the CGHS model
\cite{cghs}.  The CGHS model can also be derived as the spherical
reduction of a 4D model, but with charges.  In what follows, we will work
purely at the 2D level without any recourse to higher-dimensional physics.
The CGHS action coupled to $N$ conformal matter fields with
Lagrangian density $\mathcal{L}_m$ is
\begin{equation}
\int d^2x \sqrt{-g} \left[ e^{-2\phi} (R + 4 (\nabla \phi)^2 + 4)
+ \mathcal{L}_m \right]\, .
\end{equation}
For a null vector $k^a$, the equations of motion give
\begin{equation}
k^a k^b T_{ab} = -2 e^{-\phi} k^a k^b \nabla_a \nabla_b e^{-\phi} + 2 k^a
k^b \nabla_a e^{-\phi} \nabla_b e^{-\phi}\, .
\end{equation}
To determine the classical ``area'' of a point, we look at the coefficient
of the Ricci scalar and learn that it is proportional to $e^{-2\phi}$.  By
studying black hole thermodynamics, the constant of proportionality can be
fixed as $A_{\text{cl}} = 8 e^{-2\phi}$.

To prove the entropy bound, we start with the following
assumptions:
\begin{description}
\item[i. ] $e^{-2\phi} (s\, e^{2\phi})' \leq 2\, T_{ab}\, k^a k^b$
\item[ii. ] $s(0) \leq - \frac{1}{4} A_{\text{cl}}' (0)$
\end{description}
We will continue to use $s \equiv -k^a s_a$ and primes denoting
$d/d\lambda$.  Putting it all together, we find
\begin{eqnarray*}
s(\lambda)
&=& e^{-2\phi(\lambda)} \int_0^\lambda d\tilde{\lambda}
\left( s(\tilde{\lambda})\, e^{2\phi(\tilde{\lambda})} \right)' +
e^{-2\phi(\lambda)} s(0) e^{2\phi(0)}\\
(i) \quad
&\leq& 2\, e^{-2\phi(\lambda)} \int_0^\lambda d\tilde{\lambda}\,
e^{2\phi(\tilde{\lambda})} k^a k^b T_{ab} (\tilde{\lambda}) +
e^{-2\phi(\lambda)} s(0) e^{2\phi(0)}\\
(eom) \quad &=& - 4\, e^{-2\phi(\lambda)} \int_0^\lambda d\tilde{\lambda}
\left( e^{-\phi(\tilde{\lambda})} \right)'' e^{\phi(\tilde{\lambda})} +4\,
e^{-2\phi(\lambda)} \int_0^\lambda d\tilde{\lambda}
\left( \phi'(\tilde{\lambda}) \right)^2
+ e^{-2\phi(\lambda)} s(0) e^{2\phi(0)}\\
&=& - 4\, e^{-2\phi(\lambda)}
\left(-\phi'(\lambda) + \phi'(0) \right)
+ e^{-2\phi(\lambda)} s(0) e^{2\phi(0)}\\
(ii) \quad &\leq& - 2 \left( e^{-2\phi(\lambda)} \right)'\, .
\end{eqnarray*}
Therefore, we find the desired relation:
\begin{equation}
\int_0^1 d\lambda\, s(\lambda) \leq 2 \left( e^{-2\phi(0)} - e^{-2\phi(1)}
\right) = \frac{1}{4} \left( A_{\text{cl}}(0) -
A_{\text{cl}}(1)\right) \,
.
\end{equation}

\subsection{CGHS in Kruskal gauge}

In the previous section, we derived the CGHS entropy bound with
manifestly covariant equations of motion and entropy conditions.
However, once we add the one-loop trace anomaly, we are only able
to obtain local equations of motion in conformal gauge.
Furthermore, our calculations are greatly simplified in Kruskal
gauge.  Therefore, it behooves us to rederive the CGHS result in
Kruskal gauge.

We will assume that the lightsheet moves in the decreasing $x^+$ 
direction.
Our results for this past-directed $x^+$ lightsheet generalize simply
to the other three possible lightsheet directions.
Working with the $x^+$ lightsheet, we will be interested in the following 
equation of motion:
\begin{equation}
T_{++} = -2 e^{-\phi} \nabla_+ \nabla_+ e^{-\phi} + 2 \nabla_+
e^{-\phi} \nabla_+ e^{-\phi}\, .
\end{equation}
In conformal gauge, the RHS can be written as
$2 e^{-2\phi} \left( \partial_+ \partial_+ \phi - 2 \partial_+
\rho \partial_+ \phi\right)$.
In Kruskal gauge, we set $\rho = \phi$, so this becomes
\begin{equation}
T_{++} = -\partial_+ \partial_+ e^{-2\phi}\, .
\end{equation}

Since $k^+ = -\partial x^+ / \partial \lambda = e^{-2\phi}$ in Kruskal
gauge, our entropy conditions
can be rewritten in Kruskal gauge coordinates as
\begin{description}
\item[i.] $\partial_+ s_+  \leq 2\, T_{++}$
\item[ii.] $-s_+(x^+_0) \leq \frac{1}{4}\, \partial_+
A_{\text{cl}}(x_0^+)$.
\end{description}
Recall that $s \equiv - k^+ s_+$, so $-s_+$ is positive so long as the 
proper entropy flux $s$ is positive.

Applying these conditions, we find
\begin{eqnarray*}
-s_+(x^+) &=& \int_{x^+}^{x_0^+} d\tilde{x}^+ \partial_+
s_+(\tilde{x}^+) - s_+(x_0^+)\\
(i) \quad &\leq& 2 \int_{x^+}^{x_0^+} d\tilde{x}^+
T_{++}(\tilde{x}^+) - s_+(x_0^+)\\
(eom) \quad &=& 2 \left. \partial_+ e^{-2\phi}
\right]_{x_0^+}^{x^+} - s_+(x_0^+)\\
(ii) \quad &\leq& 2\, \partial_+ e^{-2\phi(x^+)}\, .
\end{eqnarray*}
We find that
\begin{equation}
\int_0^1 d\lambda\, s(\lambda) = \int_{x_0^+}^{x_1^+} d\tilde{x}^+
s_+(\tilde{x}^+) = \int_{x_1^+}^{x_0^+} d\tilde{x}^+ \left( 
-s_+(\tilde{x}^+) \right)
\leq \frac{1}{4} \left(
A_{\text{cl}}(0) - A_{\text{cl}}(1) \right) \, .
\end{equation}

Had we chosen the future-directed $x^+$ lightsheet, then we would have 
$k^+
= \partial x^+ / \partial \lambda = e^{-2\phi}$ and our entropy
conditions would have been $\partial_+ \left(-s_+\right)  \leq 
2\, T_{++}$
and $-s_+(x^+_0) \leq -\frac{1}{4}\, \partial_+ A_{\text{cl}}(x_0^+)$.
The extension to $x^-$ lightsheets is trivial.

\section{Stating and proving a quantum Bousso bound}

The classical CGHS action is
\begin{equation}
S_{\text{CGHS}} = \int d^2x \sqrt{-g} \left[ e^{-2\phi} (R + 4
(\nabla \phi)^2 + 4)
+ \mathcal{L}_m \right]\, .
\end{equation}

For $N$ conformal matter fields, Hawking radiation and its
backreaction on the geometry can be accounted for by adding to the
classical CGHS action the nonlocal term
\begin{equation}
S_{\text{PL}} = -\frac{N}{48} \int d^2x \sqrt{-g(x)} \int d^2x'
\sqrt{-g(x')} R(x) G(x,x')  R(x')\, ,
\end{equation}
where $G(x,x')$ is the Green's function for $\nabla^2$.
This is a one loop quantum correction but it nevertheless
contributes at leading order in a  $\frac{1}{N}$ expansion. At the
one loop level, there is the freedom of also adding a local
counterterm to the action.  The large N theory becomes analytically
soluble if we add the following judiciously chosen local counterterm to
the action
\cite{rst}:
\begin{equation}
S_{\text{ct}} = -\frac{N}{24} \int d^2 x \sqrt{-g} \phi R\, .
\end{equation}
The full action for the RST model is then
\begin{equation}
S_{\text{RST}} = S_{\text{CGHS}} + S_{\text{PL}} + S_{\text{ct}}\, .
\end{equation}

We can once again choose Kruskal gauge, but this time $\rho = \phi +
\frac{1}{2} \log (N/12)$.
In conformal and Kruskal gauges, the equations of motion become
\begin{equation}
\partial_+ \partial_- \Omega = -1\, ,
\end{equation}
and
\begin{equation}
\partial_{\pm}^2 \Omega = -\frac{12}{N} T_{\pm \pm} -t_\pm\, ,
\label{eq++eom}
\end{equation}
where
\begin{equation}
\Omega = \frac{12}{N} e^{-2\phi} + \frac{\phi}{2} + \frac{1}{4} \log
\frac{N}{48}\, .
\end{equation}

The $t_{\pm}$ term in (\ref{eq++eom}) accounts for the
normal-ordering ambiguity. We wish to consider semiclassical
excitations built on the vacuum state which has no positive
frequency excitations according to inertial observers on
$\mathcal{I}^-$. These inertial coordinates, $\sigma^{\pm}$, are
related to the Kruskal coordinates by
\begin{equation}
x^+ = e^{\sigma^+}\, , \quad x^- = -e^{-\sigma^-}\, .
\end{equation}
For coherent states built on this $\sigma$ vacuum $t_\pm=0$ in
$\sigma$ coordinates. Its value in Kruskal coordinates is given by
the Schwarzian transformation law as
\begin{equation}
t_{\pm} = -\frac{1}{4(x^{\pm})^2}\, .
\end{equation}

As worked out in \cite{fpst}, the classical ``area'' of a point in the RST
model is
\begin{equation}
A_{\text{cl}} = 8 e^{-2\phi} - \frac{N}{3} \phi
- \frac{N}{6} - \frac{N}{6} \log \frac{N}{48}\, .
\end{equation}
For coherent states built on the $\sigma$ vacuum, the renormalized
entanglement entropy across a point has a local contribution
\begin{equation}
S_{\text{ent}} = \frac{N}{6} \left(
\phi + \frac{1}{2} \log \frac{N}{12} + \frac{1}{2} \log \left( -x^+ x^-
\right) \right)\, .
\end{equation}
The full entanglement entropy also has an infrared term which is
not locally associated to the horizon and so is not included here.
See \cite{fpst} for a detailed derivation and discussion of these
points.


Now, $\Omega$ can be written as
\begin{equation}
\Omega = \frac{3}{2N} \left( A_{\text{cl}} + 4S_{\text{ent}}
\right) -\frac{1}{2} \log (-x^+ x^-) - \log 2 + \frac{1}{4}\, .
\end{equation}
Differentiating, we obtain
\begin{equation}
\partial_+ \Omega + \frac{1}{2x^+} = \frac{3}{2N} \partial_+
A_{\text{qu}}\, .
\end{equation}

When analyzing the RST model, we will leave entropy
condition \textbf{i} unchanged.  In the formulation of
\textbf{ii}, we will replace $A_{\text{cl}}$ with $A_{\text{qu}} \equiv
A_{\text{cl}} + 4 S_{\text{ent}}$
In Kruskal coordinates, these become
\begin{description}
\item[i.] $\partial_+ s_+  \leq 2\, T_{++}$
\item[ii.] $-s_+(x_0^+) \leq  \frac{1}{4} \partial_+
A_{\text{qu}} (x_0^+)$
\end{description}

\begin{eqnarray*}
-s_+(x^+) &=& \int_{x^+}^{x_0^+} d\tilde{x}^+ \partial_+ 
s_+(\tilde{x}^+) - s_+(x_0^+) \\
(i) \quad &\leq& 2  \int_{x^+}^{x_0^+}
d\tilde{x}^+ T_{++}(\tilde{x}^+) - s_+(x_0^+) \\
(eom) \quad &=& \frac{N}{6} \left. \left( \partial_+ \Omega
 + \frac{1}{4x^+} \right) \right]_{x_0^+}^{x^+} - s_+(x_0^+)\\
&=& \left.  \left( \frac{1}{4} \partial_+ A_{\text{qu}} -
\frac{N}{24x^+}
\right) \right]_{x_0^+}^{x^+} - s_+(x_0^+)\\
(ii) \quad &\leq& \frac{1}{4} \partial_+ A_{\text{qu}}(x^+) -
\frac{N}{24} \frac{1}{x^+} + \frac{N}{24} \frac{1}{x_0^+}\\
&\leq&  \frac{1}{4} \partial_+ A_{\text{qu}}(x^+)\, .
\end{eqnarray*}

We find
\begin{equation}
\int_0^1 d\lambda \, s(\lambda) = \int_{x_0^+}^{x_1^+}
d\tilde{x}^+ s_+(\tilde{x}^+) 
=\int_{x_1^+}^{x_0^+} d\tilde{x}^+ \left(-s_+(\tilde{x}^+) \right) \leq
\frac{1}{4} \left( A_{\text{qu}}(0) -
A_{\text{qu}}(1) \right)\, . \label{eqqbb}
\end{equation}
With our entropy conditions, we see that the covariant entropy bound is
satisfied once we replace $A_{\text{cl}}$ with $A_{\text{qu}}$.

It is interesting to note that the quantum Bousso bound can not be
saturated for coherent states built on the $\sigma$ vacuum.  The
obstruction to saturation is the term $\frac{N}{24} \left(\frac{1}{x_0^+} 
-\frac{1}{x^+}\right)$ that shows up in the calculation of $-s_+(x^+)$.
However, had we built our state on top of the Kruskal vacuum (i.e., the 
Hartle-Hawking state), we would
have $t_+ = 0$ and $S_{\text{ent}} = \frac{N}{6} (\phi + \frac{1}{2} \log
\frac{N}{12})$.  As a result, both our equations of motion and our
definition of $A_{\text{qu}}$ would change in a way that eliminates the
$\frac{N}{24} \left(\frac{1}{x_0^+} - \frac{1}{x^+}\right)$ term from the
calculations.  The quantum Bousso bound will then be saturated any time
the two entropy conditions are saturated.

\section*{Acknowledgements}

The work of A.S. was supported in part by DOE grant DE-FG02-91ER40654.
The work of D.T. was supported by DOE grant DE-FG02-91ER40654.  We are
grateful to R. Bousso, E. Flanagan, and A. Maloney for useful 
conversations.

\end{document}